\documentclass[11pt]{article}
\usepackage[utf8]{inputenc}
\usepackage{geometry}
\usepackage{authblk}
\usepackage{graphicx}
\usepackage[dvipsnames]{xcolor}
\definecolor{citegreen}{RGB}{40, 200, 40}
\usepackage{hyperref}
\hypersetup{colorlinks,
            linkcolor=MidnightBlue,
            urlcolor=Aquamarine,
            citecolor= citegreen,
}
\usepackage{subfig}
\usepackage{lscape}
\usepackage{sectsty}
\allsectionsfont{\sffamily\mdseries\upshape}
\usepackage[titles,subfigure]{tocloft}

\title{Cosmodoit: A Python Package for Adaptive, Efficient Pipelining of Feature Extraction from Performed Music}
\author[1]{Corentin Guichaoua}
\author[1]{Daniel Bedoya}
\author[2]{Elaine Chew}
\affil[1]{STMS Laboratoire (UMR9912) -- CNRS, IRCAM, Sorbonne Université, Ministère de la Culture, Paris 75004, France}
\affil[2]{Department of Engineering and School of Biomedical Engineering \& Imaging Sciences, King's College London, United Kingdom}
\date{September 26, 2024}

\begin{document}
\maketitle

\section{Introduction}

Computational analysis of performed music is an essential part of music information research (MIR). Almost all the music that we hear has been shaped through performance. Music performance analysis~\cite{Lerch2020} focuses on the representation and characterisation of acoustic variations performers introduce when communicating music. It is sometimes referred to as musical prosody~\cite{Palmer2006}. These acoustic variations are shaped by individual interpretations of music pieces. They encode performers' conceptualisations of music structures~\cite{Guichaoua2024} and provide valuable cues for listeners' response to music~\cite{Cotic2024}.

Music performance analysis has grown considerably in recent years with the development of algorithms and tools in programming languages such as Matlab (\cite{Pampalk2004,Devaney2014}), C++ (\cite{Nakamura2017}), and Python (\cite{Guo2019}\linebreak\cite{Peter2023}). Each algorithm or tool is tailored to perform a specific task out of several that are critical to the analysis of performed music. The diversity of implementation languages and data formats makes existing algorithms difficult to combine. There is an urgent need for a comprehensive and efficient tool that can streamline the process of feature extraction from performed music. This is the aim of Cosmodoit.

Cosmodoit is a novel Python package that combines performance-to-score alignment and symbolic and audio music feature extraction. The software is designed to be an adaptive and efficient tool that can process performed music at scale. The pipeline of Cosmodoit is designed to be flexible, allowing users to perform selective processing and compute only the desired features and the ones on which they depend, thus avoiding duplicated work and reducing computational time.

The design of Cosmodoit was based on several key considerations, including the requirement to simplify the process of music performance analysis, reducing errors caused by file changes or swaps, and providing a tool that can be updated regularly to accommodate evolving needs. The package is modular and extensible, allowing users to add or ignore features as needed. In addition, Cosmodoit allows for the computation of small increments (like adding features) as required, making it an ideal tool for both research and development purposes.

One of the key features of Cosmodoit is its ability to handle algorithms from a variety of programming languages, making it a versatile tool for music performance analysis. The package also has the ability to change parameters such as window length, allowing users to fine-tune the pipeline and extract the desired features from performed music while ensuring the consistency for all dependent features. This makes Cosmodoit a handy tool for researchers and developers in the field of music performance analysis.

\section{Statement of need}
Data collection in music performance can be a very slow process, involving inviting performers, recording sessions and gathering additional data such a scores. As a result, new research on methods and insights is often intertwined with building datasets, which leads to a growing need for re-computing features because of new or updated data and changes to the downstream code and its parameters.

The process of combining features in music performance analysis has traditionally been a manual one, requiring researchers and developers to keep track of the order in which features are extracted and combined. This manual process can be time-consuming and error-prone, as updates to the data or the code can require that the entire pipeline be re-run or risk missing a dependency resulting in out-of-sync features. This is compounded by the variety of environments, as some steps may require running processes in proprietary software, while others are long calls from the command line. As software in the analysis pipeline are usually developed independently, the tools being combined in sequence often do not share common input and output formats, which introduces yet an extra step where errors can occur.

\section{Design Choices}
Python is a widely-used programming language in music performance analysis, and many existing algorithms and tools are available in this language. Building the pipeline in Python makes it easier for researchers and developers to integrate their existing code into the Cosmodoit framework, while enabling code wrapping from most other languages as well.

Cosmodoit was built using Doit~\cite{Schettino2018}, a Python-based build system, as its foundation. Doit has a proven track record of being a reliable and efficient build system, having been used successfully in a variety of projects not limited to compiling code. On top of the dependency and up-to-date tracking of Doit, Cosmodoit dynamically handles the discovery of newly added performances and of which tasks can be run base on available file types---e.g., skipping the score alignment when no score is available---such that updating the features can usually be done in a single-word command. 
The modular structure of the package also allows for selective processing, where the user can choose to process only specific features, reducing the computational time. This not only saves time, but also allows for efficient exploration of evolving methods and datasets.

\begin{landscape}
    \begin{figure}
    \centering
    \includegraphics[width=1.5\textwidth]
    {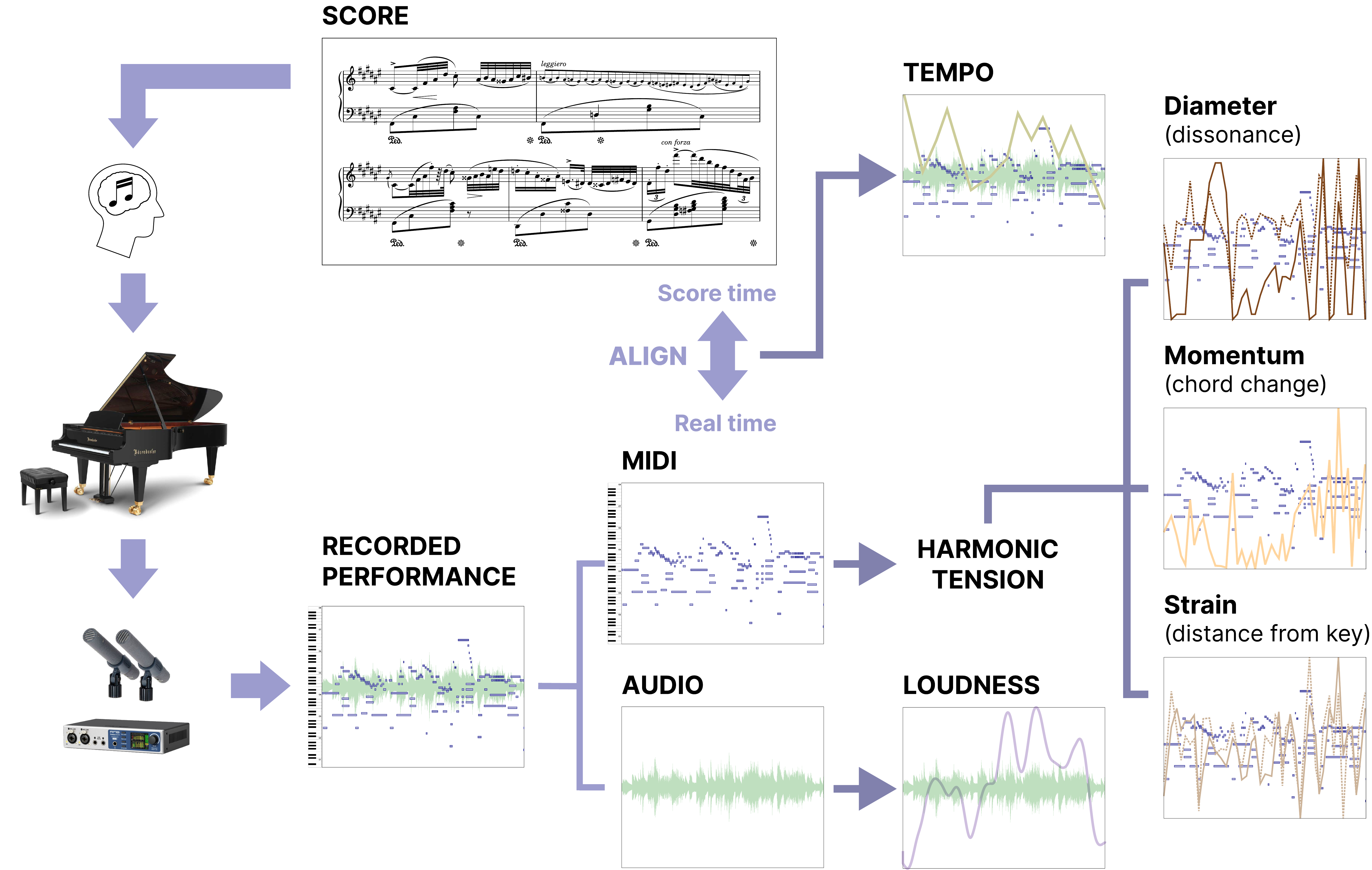}
    \caption{Cosmodoit system processes initiated by a single command-line call. Example based on Raoul Pugno's performance of Chopin's Nocturne Op.15 No.2 in F$\sharp$ major (Legendary Artists Collection courtesy of Bösendorfer). Cosmodoit's output data streams are visualised on the web based citizen science platform CosmoNote~\cite{Fyfe2022}.}
    \label{fig:system}
    \end{figure}
\end{landscape}

Cosmodoit currently contains three main submodules based on prior work: a new Python port of the loudness computation from Elias Pampalk's Matlab Music Analysis Toolbox~\cite{Pampalk2004}, a wrapper module for Eita Nakamura's C++ MIDI-to-MIDI music alignment~\cite{Nakamura2017}, and Rui Guo's Python implementation of \linebreak \cite{Herremans2016}'s Java-based harmonic tension computations in midi-miner \linebreak \cite{Guo2019}. It also includes a basic beat finding algorithm based on the result of the alignment and small modules to extract note velocity and sustain pedal values from a performed MIDI file. Figure~\ref{fig:system} shows a typical set of processes triggered by a single command-line call.

Where relevant, such as for the loudness computation, modules can expose parameters of the algorithm through a configuration file to override their default values at the dataset level
, without risk of naming clashes. A template of such a file is provided in the repository, with all exposed parameters for the existing modules. Changing parameters will trigger the re-computation of affected features on the next run. While the current version does not contain alternative algorithms for any given feature, the parameter system can be leveraged to switch between different methods. 

Finally, thanks to the modular structure of the package, new modules to compute additional features can easily be added with minimal modification of the \texttt{dodo.py} file, which serves as the main entry point. Provided the new module follows the template of existing modules, the corresponding tasks will be incorporated into the pipeline and have their dependencies detected and ran automatically.

\section{Acknowledgements}

This result is part of the project COSMOS that has received funding from the European Research Council under the European Union's Horizon 2020 research and innovation program (Grant agreement No. 788960).

\end{document}